\documentclass[prl,twocolumn,superscriptaddress]{revtex4}
\usepackage[ansinew]{inputenc}
\usepackage[dvips]{graphicx}
\usepackage{amsmath}
\usepackage{amsthm}
\usepackage{layout}
\usepackage{float}
\usepackage{subfigure}
\usepackage{amsfonts}
\usepackage{amssymb}

\begin{document}

\title{Experimental Entangled Entanglement}
\author{Philip Walther$^{1\ast}$, Kevin J. Resch$^{1\dagger}$, \v{C}aslav
Brukner$^{1}$ and Anton Zeilinger$^{1,2}$}
\affiliation{$^{1}$Institut f\"{u}r Experimentalphysik, Universit\"{a}t Wien,
Boltzmanngasse 5, A-1090 Wien, Austria \\
$^{2}$IQOQI, Institut f\"{u}r Quantenoptik und Quanteninformation, \"{O}%
sterreichische Akademie der Wissenschaften, Boltzmanngasse 3,
A-1090 Wien, Austria}

\begin{abstract}
All previous tests of local realism have studied correlations between
single-particle measurements. In the present experiment, we have performed a
Bell experiment on three particles in which one of the measurements
corresponds to a projection onto a maximally-entangled state. We show
theoretically and experimentally, that correlations between these entangled
measurements and single-particle measurements are too strong for any
local-realistic theory and are experimentally exploited to violate a
CHSH-Bell inequality by more than 5 standard deviations. We refer to this
possibility as ``entangled entanglement''.
\end{abstract}

\maketitle

Seventy years ago Einstein, Podolsky and Rosen (EPR) argued that quantum
theory~could not be a complete description of physical reality, based firmly
on plausible assumptions about locality, realism, and theoretical
completeness \cite{13}. It was not until almost 30 years later that the EPR
paradox was formulated in terms of an experimentally-testable prediction,
discovered by John Bell~\cite{1}, where the assumptions of locality and
realism put measurable limits on the strength of correlations between
outcomes of remote measurements. Since Bell's discovery, these limits, known
as Bell's inequalities, have been subject to a large number and diverse
range of experimental tests~\cite{2}. All previous Bell experiments measure
degrees of freedom corresponding to properties of \textit{individual}
systems. In these Bell experiments the joint properties of two or more
particles, which correspond to the specific type of their entanglement,
could still be independent of the measurements performed.

Bell's argument can be applied to outcomes of any measurements. In the
present work, we experimentally demonstrate the first example of a Bell
inequality test in which one of the measurements is a projection onto a
maximally entangled state. The measurement on a single particle by ``Alice''
defines a relational property between another two particles without defining
their single-particle properties. These relational properties are measured
by ``Bob''. Correlations between the measurement outcomes of the
polarization state of a single photon and the entangled state of another two
are experimentally demonstrated to violate the CHSH-Bell inequality~\cite{12}%
. This shows that \textit{entanglement itself can be entangled}~\cite{100}.

We begin with a brief discussion of two-qubit entanglement. Consider the
state, $|\phi ^{-}\rangle _{a,b}=\frac{1}{\sqrt{2}}(|H\rangle _{a}|H\rangle
_{b}-|V\rangle _{a}|V\rangle _{b}),$which is one of the four Bell states, $%
|\phi ^{\pm }\rangle _{a,b}=\frac{1}{\sqrt{2}}(|H\rangle _{a}|H\rangle
_{b}\pm |V\rangle _{a}|V\rangle _{b})$ and $|\psi ^{\pm }\rangle _{a,b}=%
\frac{1}{\sqrt{2}}(|H\rangle _{a}|V\rangle _{b}\pm |V\rangle _{a}|H\rangle
_{b})$. The subscripts $a$\ and $b$\ label Alice and Bob's photons and the
kets $|H\rangle $\ and $|V\rangle $\ and represent states of horizontal and
vertical polarization. The entangled state $|\phi ^{-}\rangle _{a,b}$ is
special in that the individual photons have perfect polarization
correlations at any angle in the $y$-$z$ plane of the Bloch sphere. It is
well known that any Bell state is capable of violating the CHSH Bell
inequality at Cirel'son's bound~\cite{14}.

In our experiment, we need a quantum state in which the polarization state
of one photon is non-classically correlated to the entangled state of the
other two in a manner that is directly analogous to that in $|\phi
^{-}\rangle _{a,b}$. An example of such a state can be written down by
replacing the polarization state of particle $b$ with the Bell states, $%
|\phi ^{-}\rangle _{b_1,b_2}$ and $|\psi ^{+}\rangle _{b_1,b_2}$,
resulting in
\begin{equation}
|\Phi ^{-}\rangle _{a,b}=\frac{1}{\sqrt{2}}(|H\rangle _{a}|\phi
^{-}\rangle _{b_1,b_2}-|V\rangle _{a}|\psi ^{+}\rangle
_{b_1,b_2}),
\end{equation}%
where now Bob possesses two photons instead of just one. Drawing on our
understanding of the bipartite entanglement, we can immediately make the
following statements. Firstly, the perfect correlations between polarization
of Alice's photon and joint properties of Bob's two photons imply, by EPR
premises of locality and realism~\cite{13}, that the entangled states of
Bob's photons are elements of reality. Secondly, each of Bob's two photons
have no well-defined individual properties, i.e., individual detection
events at Bob's detectors are random and cannot be inferred by a linear
polarization measurement by Alice. Therefore, the entangled state that Bob
possesses is an element of physical reality in the EPR sense whereas his
individual photons are not.

Interestingly, the state in Eq. 1 is equivalent to the
Greenberger-Horne-Zeilinger (GHZ) state~\cite{15}, $|\Phi ^{-}\rangle _{a,b}=%
\frac{1}{\sqrt{2}}(|R\rangle _{a}|R\rangle _{b_1}|R\rangle
_{b_2}+|L\rangle
_{a}|L\rangle _{b_1}|L\rangle _{b_2})$, where $|R/L\rangle =1/\sqrt{2}%
(|H\rangle \pm i|V\rangle )$ represent the different senses of
circularly-polarized light and the subscripts $b1$ and $b2$ each
photon at Bob's side. This connection suggests practical
preparation methods, since three-qubit GHZ states have been
generated~\cite{16}. While such GHZ states have been used in
tests of local realism through Mermin, Ardehali and
Klyshko~\cite{18} inequalities or quantum erasers~\cite{19}, we
stress that these experiments are based solely on the
correlations between single-particle observables. Quantum state
tomography~\cite{james} has been used to reconstruct quantum
states of up to 4 photons~\cite{cluster} and up to 8
ions~\cite{ions}; again these experiments are based on
measurements of correlations between single-particle observables.
No previous experiment has directly measured correlations in
violation of local realism using more complex two-particle
observables.

In Fig. 1 we show a schematic for our experiment in which a
source emits an entangled state of three particles. Alice
receives a single photon and Bob receives the other two. Alice
chooses a measurement setting in the form of an angle,
$\theta_{1}$. She then makes linear polarization measurements
parallel to that angle or perpendicular to it by orienting a
linear polarizer. A value of +1(-1) is assigned to those outcomes
where the photon is measured with the linear polarizer parallel
(perpendicular) to $\theta_{1}$. Similarly, Bob performs a
restricted Bell-state measurement in the subspace , $|\phi ^{-}\rangle _{b_1,b_2}$ or $%
|\psi ^{+}\rangle _{b_1,b_2}$. These two Bell states can be
coherently mixed using the half-wave plate (HWP)~\cite{22} in
mode b2 so that Bob's Bell state analyzer makes projective
measurements onto the maximally-entangled state $\cos \theta
_{2}|\phi ^{-}\rangle _{b_1,b_2}+\sin \theta _{2}|\psi ^{+}\rangle
_{b_1,b_2}$. By analogy with the polarization measurements, Bob
assigns a value of +1 or -1 for measurements when the HWP is set
to $\theta_2 /2$ or $(\theta_2+\pi/2)/2$, respectively. For
consistency throughout this paper we have adopted the convention
for the angle $\theta$ to mean the rotation of a polarization in
real space. Therefore the same polarization rotation on the Bloch
sphere is $2\theta $ and that rotation is induced by a HWP which
is itself rotated by $\theta/2$ . When Alice and Bob choose the
orientations $\theta _{1}$ and $\theta _{2}$ their shared
entangled state transforms to
\begin{eqnarray}
|\Phi ^{-}\rangle _{a,b} &=&\cos (\theta _{1}+\theta _{2})\frac{1}{\sqrt{2}}%
(|H\rangle _{a}|\phi ^{-}\rangle _{b_1,b_2}-|V\rangle _{a}|\psi
^{+}\rangle _{b_1,b_2})
\nonumber \\
&+&\sin (\theta _{1}+\theta _{2})\frac{1}{\sqrt{2}}(|V\rangle
_{a}|\phi ^{-}\rangle _{b_1,b_2}+|H\rangle _{a}|\psi ^{+}\rangle
_{b_1,b_2}),
\end{eqnarray}%
which entails perfect correlations for any local settings $\theta _{1}$ and $%
\theta _{2}$ such that $\theta _{1}+\theta _{2}=0,$ $\pi ,$ $2\pi ,$ $etc.$
and perfect anti-correlations when $\theta _{1}+\theta _{2}=\pi /2,$ $3\pi
/2,$ $etc.$ Imperfectly correlated and anti-correlated events will occur at
angles away from these specific settings and form the basis of a test of
local realism.

We generate our three-photon state using a pulsed ultraviolet laser (pulse
duration 200 fs, repetition rate 76 MHz) which makes two passes through a
type-II phase-matched $\beta $-barium borate (BBO) nonlinear crystal~\cite%
{23}, in such a way that it emits highly polarization-entangled photon pairs
into the modes $a1$ and $b1$ and $a2$ and $b2$ (Fig. 2). Transverse and
longitudinal walk-off effects are compensated using a HWP and an extra BBO
crystal in each mode. By additionally rotating the polarization of one
photon in each pair with additional HWPs and tilting the compensation
crystals, any of the four Bell states can be produced in the forward and
backward direction. We align the source to produce the Bell state, $|\phi
^{+}\rangle $, on each pass of the pump. Photons are detected using
fibre-coupled single-photon counting modules. We spectrally and spatially
filter the photons using 3-nm bandwidth filters and single-mode optical
fibers. We measured 26000 polarization-entangled pairs into the modes $a1$
and $b1$ and 18000 pairs into the modes $a2$ and $b2$. The visibilities of
each pair were measured to exceed $95\%$ in the $|H/V\rangle $ basis and $%
94\%$ in the $|\pm \rangle =1/\sqrt{2}(|H\rangle \pm |V\rangle )$ basis. \
Parametric down-conversion is a probabilistic emitter of photon pairs and as
such can sometimes emit two pairs of photons from the same pump pulse into
the same pair of modes. In our experiment, four-fold coincidence events from
double-pair emission is highly suppressed by a quantum interference effect
due to the polarization rotation incurred in the quarter-wave plate (QWP)
and the polarizing beamsplitter (PBS) \cite{24}.

To generate our target state, we superpose one photon from each pair, those
in modes $a1$ and $a2$, on PBS1. The PBS implements a two-qubit parity check~%
\cite{26}: if two photons enter the PBS from different input
ports, then they must have the same polarization in the
$|H/V\rangle $ basis in order to pass to the two different output
ports. Provided the photons overlap at the PBS, the initial
state, $|\phi ^{+}\rangle _{a_1,b_1}|\phi ^{+}\rangle
_{a_2,b_2}$, is converted to the GHZ\ state $|\Psi \rangle =\frac{1}{\sqrt{2}}%
(|H\rangle _{T}|H\rangle _{a}|H\rangle _{b_1}|H\rangle
_{b_2}+|V\rangle _{T}|V\rangle _{a}|V\rangle _{b_1}|V\rangle
_{b_2})$ provided the photons emerge into different output spatial
modes~\cite{27}. Rotations incurred in
QWPs and the subsequent projection of the trigger photon in mode $T$ onto $%
|H\rangle _{T}$ reduces the four-particle GHZ state to the desired
three-photon entangled entangled state $|\Phi ^{-}\rangle _{a,b}$.

The polarization of Alice's photon was measured with a polarizer
oriented along the angle $\theta _{1}$. Bob's measurements were
made using a Bell-state analyzer based on a PBS~\cite{28}. By
performing a check that the parity of Bob's photons is even, the
PBS acts as a $|\phi ^{\pm }\rangle _{b_1,b_2} $-subspace filter.
The two Bell states in this subspace, $|\phi ^{+}\rangle
_{b_1,b_2}$ and $|\phi ^{-}\rangle _{b_1,b_2}$, have opposite
correlations in the $|\pm \rangle $ basis and are distinguished
using linear polarizers.
One polarizer oriented along the $|+\rangle $ direction and the other along $%
|-\rangle $ completes a projective measurement onto $|\phi ^{-}\rangle _{b_1,b_2}$%
. The setting of the HWP in mode $b2$ before PBS2 allows
projection any state of the form $\cos \theta _{2}|\phi
^{-}\rangle _{b_1,b_2}+\sin \theta _{2}|\psi ^{+}\rangle
_{b_1,b_2}$.

Correlation measurements between Alice and Bob were made by rotating Alice's
polarizer in $30^{\circ }$ steps while Bob's HWP was kept fixed at $\theta
_{2}/2=0^{\circ }$ or $22.5^{\circ }$. Four-fold coincidence counts at each
setting were measured for 1800 seconds (Fig. 3). The count rates follow the
expected relation $N(\theta _{1},\theta _{2})\propto \cos ^{2}(\theta
_{1}+\theta _{2})$ with visibilities of $(78\pm 2)\%$ in the $|H/V\rangle $
basis and $(83\pm 2)\%$ in the $|\pm \rangle $ basis. Both surpass the
crucial limit of \symbol{126}$71\%$ which, in the presence of white noise,
is the threshold for demonstrating a violation of the CHSH-Bell inequality.

For our state, $|\Phi ^{-}\rangle _{a,b}$, the expectation value
for the correlations between a polarization measurement at Alice
and a maximally-entangled state measurement at Bob is $E(\theta
_{1},\theta _{2})=\cos [2(\theta _{1}+\theta _{2})]$. The
correlation can be expressed
in terms of experimentally-measurable counting rates using the relation $%
E(\theta _{1},\theta _{2})=\frac{N^{++}+N^{--}-N^{+-}-N^{-+}}{%
N^{++}+N^{--}+N^{+-}+N^{-+}}$, where $N$ is the number of coincidence
detection events between Alice and Bob with respect to their set of analyzer
angles $\theta _{1}$ and $\theta _{2}$, where $+1(-1)$ outcomes are denoted
as ``$+$''(``$-$''). These correlations can be combined to give the
CHSH-Bell parameter, $S=\left| -E_{1}(\theta _{1},\theta _{2})+E_{2}(\tilde{%
\theta _{1}},\theta _{2})+E_{3}(\theta _{1},\tilde{\theta _{2}})+E_{4}(%
\tilde{\theta _{1}},\tilde{\theta _{2}})\right| $, which is maximized at $%
\{\theta _{1},\tilde{\theta _{1}},\theta _{2},\tilde{\theta _{2}}%
\}=\{0^{\circ },45^{\circ },22.5^{\circ },67.5^{\circ }\}$ to $S=2\sqrt{2}$
. This violates the inequality $S\leq 2$ for local realistic theories.

In Fig. 4, the count rates for the 16 required measurement settings to
perform the CHSH inequality are shown and give the four correlations $%
E_{1}(\theta _{1},\theta _{2})=0.69\pm 0.05$, $E_{2}(\tilde{\theta _{1}}%
,\theta _{2})=-0.61\pm 0.04$ , $E_{3}(\theta _{1},\tilde{\theta _{2}}%
)=-0.58\pm 0.04$, and $E_{4}(\tilde{\theta _{1}},\tilde{\theta _{2}}%
)=-0.60\pm 0.04$. These correlations yield the Bell parameter, $S=2.48\pm
0.09$ which strongly violates the CHSH-Bell inequality by 5.6 standard
deviations.

We have demonstrated a violation of the CHSH-Bell inequality using the
correlations between a single particle property, the polarization state of a
photon, and a joint property of two particles, the entangled state of a
photon pair. In doing so we have experimentally demonstrated that
two-particle correlations have the same ontological status as
single-particle properties. Our result shows that it only makes sense to
speak about measurement events (detector ``clicks'') whose statistical
correlations may violate limitations imposed by local realism and thus
indicate entanglement.

The authors thank M. Aspelmeyer for helpful discussions. This work was
supported by the Austrian Science Foundation (FWF), NSERC, the European
Commission, Marie Curie Fellowship, and under project RAMBOQ.

Fig. 1. Schematic cartoon for the Bell experiment based on an
entangled entangled state. a) A source emits the entangled
three-photon state, $|\Phi ^{-}\rangle
_{a,b}=\frac{1}{\sqrt{2}}\left( |H\rangle _{a}|\phi ^{-}\rangle
_{b_1,b_2}-|V\rangle _{a}|\psi ^{+}\rangle _{b_1,b_2}\right) $,
where one photon is received by Alice and the two other photons
by Bob. Alice makes polarization measurements on her photon. If
the photon's polarization is measured to be parallel
(perpendicular) to the orientation, $\theta _{1}$, of the
analyzer, the measurement outcome is $+1$ $(-1)$. In contrast,
Bob makes projective measurements onto a two-particle entangled
state, where the orientation of his analyzer is defined by the
angle, $\theta _{2}$. Bob's outcomes are also defined as $+1$ or
$-1$. b) For the local settings $\theta _{1}+\theta _{2}=0,$ $\pi
,$ $2\pi ,$ etc. they observe perfect correlations, i.e., the
product of their local measurement outcomes yields $+1$. c)
Perfect
anti-correlations will be obtained when $\theta _{1}+\theta _{2}=\pi /2,$ $%
3\pi /2,$ etc., given by product $-1$ of the local results.

Fig. 2. Setup for the experimental realization. A spontaneous
parametric down-conversion source emits $|\phi ^{+}\rangle $
states, into both pair of modes $a1$ \& $b1$ and $a2$ \& $b2$.
Comp is a 1mm thick BBO crystal used to compensate the walkoff in
the down-conversion crystal~\cite{23}. The modes $a1$ and $a2$
are superposed at the polarizing beamsplitter PBS1. In our case
the PBS transmits
horizontally-polarized and reflects vertically-polarized photons. Each mode $%
T$, $a$, $b1$, $b2$ passes through a quarter-wave plate (QWP).
Projecting the trigger qubit $T$ onto the state $|H\rangle _{T}$,
we generate the state $|\Phi ^{-}\rangle
_{a,b}=\frac{1}{\sqrt{2}}(|H\rangle _{a}|\phi ^{-}\rangle
_{b_1,b_2}-|V\rangle _{a}|\psi ^{+}\rangle _{b_1,b_2})$. The
photon in mode $a$ is sent to Alice, who makes single-photon
polarization measurements, determined by the orientation angle,
$\theta _{1}$, of her linear polarizer. The photons in mode $b1$
and $b2$ belong to Bob, who uses a modified Bell-state analyzer
to make projective measurements onto a coherent superposition of
$|\phi
^{-}\rangle _{b_1,b_2}$ and $|\psi ^{+}\rangle _{b_1,b_2}$, where the mixing angle, $%
\theta _{2}$, is determined by the angle $\theta _{2}/2$, of the half-wave
plate (HWP) in mode $b2$.

Fig. 3. Measured coincidence fringes for the entangled entangled
state. Bob's half-wave plate was initially set to $0^{\circ }$
and made projective measurements onto the state $|\phi
^{-}\rangle _{b_1,b_2}$. The total number of four-fold coincidence
counts measured in 1800 seconds as a function of the angle of
Alice's polarizer is shown as solid squares. Fitting the curve to
a sinusoid (solid line) yields a visibility of $(78\pm 2)\%$.
After changing
Bob's measurement setting to project onto the state $\frac{1}{\sqrt{2}}%
(|\phi ^{-}\rangle _{b_1,b_2}+|\psi ^{+}\rangle _{b_1,b_2})$, the
procedure was repeated. The data for these settings are shown as
open circles. The sinusoidal fit (dotted line) yields a
visibility of $(83\pm 2)\%$.

Fig. 4. Experimentally measured coincidence counting rates used to test the
CHSH-Bell inequality. The requisite coincidence measurements for the 16
different measurement settings are shown. Each measurement was performed for
1800 seconds. For measurement settings, ${\theta _{1},}$ ${\theta _{2}}$,
the axis labels $++$, $+-$, $-+$, and $--$ refer to the actual settings of $%
\left( {\theta _{1},\theta _{1}}\right) $, $\left( {\theta _{1},\theta
_{2}+\pi /2}\right) $, $\left( {\theta _{1}+\pi /2,\theta _{2}}\right) $,
and $\left( {\theta _{1}+\pi /2,\theta _{2}+\pi /2}\right) ,$ respectively.
These data yielded the Bell parameter $S=2.48\pm 0.09$ which is in conflict
with local realism by over 5 standard deviations.

\end{document}